\documentclass[12pt]{article}
\usepackage{amssymb,amsmath,cite}
\catcode`\@=11

\catcode`\@=11
\def\thesection{\arabic{section}.}

\def\appendix{\setcounter{section}{0}
        \def\thesection{Appendix.}
        \def\theequation{\Alph{section}.\arabic{equation}}}
\def\section{\@startsection{section}{1}{\z@}{3.5ex plus 1ex minus
   .2ex}{2.3ex plus .2ex}{\large\bf}}
     
\newcommand{\beq}{\begin{equation}}
\newcommand{\eeq}{\end{equation}}

\begin{document}
\thispagestyle{empty}
\vspace{.5in}
\begin{flushright}
UCD-05-01\\
gr-qc/0501033\\
January 2005\\
revised April 2005\\
\end{flushright}
\vspace{.5in}
\begin{center}
{\Large\bf
 Dynamics of Asymptotic Diffeomorphisms\\[1ex]
 in (2+1)-Dimensional Gravity}
\end{center}
\vspace*{3ex}
\begin{center}
{S.~C{\sc arlip}\footnote{\it email: carlip@physics.ucdavis.edu}\\
       {\small\it Department of Physics}\\
       {\small\it University of California}\\
       {\small\it Davis, CA 95616}\\{\small\it USA}}
\end{center}
\vspace*{3ex}
\begin{center}
{\large\bf Abstract}
\end{center}
\begin{center}
\begin{minipage}{4.75in}
{\small
In asymptotically anti-de Sitter gravity, diffeomorphisms that
change the conformal boundary data can be promoted to genuine 
physical degrees of freedom.  I show that in 2+1 dimensions, 
the dynamics of these degrees of freedom is described by a 
Liouville action, with the correct central charge to reproduce 
the entropy of the BTZ black hole.
}
\end{minipage}
\end{center}
 
\addtocounter{footnote}{-1}
\addtocounter{page}{-1}
\newpage

For those who believe that black hole entropy counts microscopic
quantum states, the (2+1)-dimensional black hole of Ba{\~n}ados,
Teitelboim, and Zanelli \cite{BTZ} presents an interesting conundrum.  
In three spacetime dimensions, general relativity is a ``topological''
theory, with only a finite, and typically small, number of global
degrees of freedom \cite{Carlipa}.  The BTZ black hole, on the other
hand, can have an arbitrarily large entropy.  Where do these states
come from?

In the Chern-Simons formulation of (2+1)-dimensional anti-de Sitter 
gravity \cite{Achucarro,Wittena}, an answer to this question is known.
Chern-Simons theory is a gauge theory, with gauge transformations
parametrized by group elements $g\in G$.  In general, fields that
differ by gauge transformations are physically identical.  But at 
a boundary---including the conformal boundary of an asymptotically 
anti-de Sitter space---gauge transformations become symmetries rather
than invariances, and gauge-equivalent fields are physically distinct.
Such inequivalent fields can be labeled by the group element $g$,
which itself becomes a dynamical field at the boundary \cite{EMSS,%
Carlipb}.  In general, the induced dynamics is that of a chiral WZW
model, but for (2+1)-dimensional gravity, slightly stronger anti-de 
Sitter boundary conditions reduce the action to that of Liouville theory
\cite{CHvD,Roomana}.  The degrees of freedom of the BTZ black hole may 
thus be viewed as ``would-be pure gauge'' degrees of freedom that become
dynamical at the boundary \cite{Carlipc}.  As a partial confirmation 
of this picture, the central charge of the induced Liouville theory%
---which matches that of the asymptotic symmetry group in the metric 
formalism \cite{BrownHenn}---has precisely the right value to reproduce,
via Cardy's formula, the correct BTZ black hole entropy \cite{Strominger}.

Unfortunately, the Chern-Simons derivations \cite{CHvD,Roomana} of 
Liouville theory rely heavily on features peculiar to 2+1 dimensions, 
and their implications for higher-dimensional gravity remain unclear.  
Liouville theory has appeared in the metric formalism in several contexts 
as well \cite{Navarro,Nakatsu,SkenSolo,Bautier,Roomanb,Krasnova,Krasnovb,%
Manvelyan}, but the Liouville field has largely been treated as an 
auxiliary field or as a field ``dual'' to the bulk degrees of freedom.  
This contrasts sharply with the Chern-Simons picture, in which the Liouville 
field is a piece of the ordinary gauge field, albeit a piece that becomes 
physical only at the conformal boundary.  

But the metric formalism also 
has a gauge-like symmetry, diffeomorphism invariance, which should also lead 
to new physical degrees of freedom at a boundary.  The goal of this paper is 
to show explicitly that these ``would-be diffeomorphism'' degrees of freedom 
are, indeed, described by a Liouville theory with the correct central charge.

We begin with a Fefferman-Graham-type expansion \cite{FG} of the metric
near infinity,
\beq
ds^2 = -\ell^2d\rho^2 + g_{ij}dx^idx^j , \qquad\hbox{with}\quad
  g_{ij} = e^{2\rho}\overset{(0)}{g}_{ij}(x) + \overset{(2)}{g}_{ij}(x)
  + \dots ,
\label{a1}
\eeq
where $i=0,1$ and the cosmological constant is $\Lambda=-1/\ell^2$.
The Einstein field equations then yield \cite{HennSken}
\beq
\overset{(2)}g\,{}^i{}_i = -\frac{\ell^2}{2}\overset{(0)}R ,\qquad
\overset{(0)}\nabla_i\overset{(2)}g_{jk} -
 \overset{(0)}\nabla_j\overset{(2)}g_{ik} = 0 ,
\label{a2}
\eeq
where indices are raised and lowered and covariant derivatives defined with 
respect to the conformal boundary metric $\overset{(0)}g_{ij}$.  It should
be noted that in 2+1 dimensions, these results do not require the full field
equations, but follow from the Hamiltonian and momentum constraints at constant
$\rho$.  As in the Chern-Simons approach, we will impose the constraints, but
need not assume the full ``bulk'' equations of motion.

Let us now consider a coordinate transformation
\begin{align}
\rho &\rightarrow \rho + \frac{1}{2}\varphi(x) + e^{-2\rho}\overset{(2)}f(x) 
  + \dots \nonumber\\
x^i &\rightarrow x^i + e^{-2\rho}\overset{(2)}h{}^i(x) + \dots ,
\label{a3}
\end{align}
and determine $\overset{(2)}f$ and $\overset{(2)}h{}^i$ by demanding that 
the metric remain in the form (\ref{a1}).  It is easy to check that
\beq
g_{\rho i} = -\frac{\ell^2}{2}\partial_i\varphi 
  - 2e^\varphi\overset{(2)}h_i , \qquad
g_{\rho\rho} = -\ell^2\left(1-4\overset{(2)}fe^{-2\rho}\right)
  + 4e^{-2\rho}e^\varphi\overset{(0)}g_{ij}\,\overset{(2)}h{}^i\overset{(2)}h{}^j
\label{a4}
\eeq
and thus \cite{Skenderis}
\beq
\overset{(2)}h_i = -\frac{\ell^2}{4}e^{-\varphi}\partial_i\varphi , \qquad
\overset{(2)}f = -\frac{\ell^2}{16}e^{-\varphi}\,\overset{(0)}g{}^{ij}
  \partial_i\varphi\partial_j\varphi .
\label{a5}
\eeq
The spatial metric in the new coordinate system may then be calculated;
one finds 
\beq
g_{ij} = e^{2\rho}e^\varphi\overset{(0)}g_{ij} + 8\pi G\ell T_{ij}
  + \left(\overset{(2)}g_{ij} 
  - \frac{\ell^2}{2}\overset{(0)}g_{ij}\overset{(0)}{\Delta}\varphi
  - \frac{\ell^2}{4}\lambda\overset{(0)}g_{ij}e^\varphi \right) + \dots
\label{a6}
\eeq
where 
\beq
T_{ij} = \frac{\ell}{32\pi G}\left[ \partial_i\varphi\partial_j\varphi
  - \frac{1}{2}\overset{(0)}g_{ij}\overset{(0)}g{}^{kl}
  \partial_k\varphi\partial_l\varphi 
  - 2\overset{(0)}\nabla_i\overset{(0)}\nabla_j\varphi 
  + 2\overset{(0)}g_{ij}\overset{(0)}{\Delta}\varphi 
  + \lambda\overset{(0)}g_{ij}e^\varphi \right]
\label{a7}
\eeq
is the stress-energy tensor for the Liouville action  
\beq
I_{\mathit Liou} = -\frac{\ell}{32\pi G}\int d^2x\sqrt{\overset{(0)}g}
  \left( \frac{1}{2}\overset{(0)}g{}^{ij}\partial_i\varphi\partial_j\varphi
  -\varphi\overset{(0)}R -\lambda e^\varphi\right) .
\label{a8}
\eeq

This appearance of the Liouville stress-energy tensor is not new: it
was introduced in \cite{SkenSolo,Bautier} as a means of integrating the 
Einstein equations (\ref{a2}), and was derived from coordinate transformations 
in \cite{Navarro,Roomanb,Krasnovb}.  What has been missing so far is the 
recognition of the field $\varphi$ in the stress-energy tensor as a genuine 
dynamical degree of freedom.

To obtain the dynamics of the Liouville field, let us start with the 
Einstein-Hilbert action
\beq
I_{\mathit grav} = \frac{1}{16\pi G}\int_M d^3x\sqrt{{}^{(3)}g}\left( 
  {}^{(3)}R + \frac{2}{\ell^2}\right) 
  + \frac{1}{8\pi G}\int_{\partial M} d^2x\sqrt{\gamma}K
  - \frac{1}{8\pi G\ell}\int_{\partial M} d^2x\sqrt{\gamma} ,
\label{a9}
\eeq
where $\gamma_{ij}$ is the induced metric on the boundary $\partial M$.  The 
last term in (\ref{a9}) is the regulator introduced in \cite{HennSken,BalKraus} 
to make the action finite in the limit that the boundary goes to infinity.  
Initially we evaluate the action in asymptotically anti-de Sitter space with 
a boundary at $\rho=\bar\rho$, with $\bar\rho\rightarrow\infty$.  With the 
coordinate transformation (\ref{a3}), though, we should place the boundary 
at a location at which the \emph{new} radial coordinate is constant; that is, 
in the original coordinate system,
\beq
\rho = \bar\rho + \frac{1}{2}\varphi + e^{-2\bar\rho}\overset{(2)}f 
  + \dots = F(x) .
\label{a10}
\eeq

The induced metric and unit normal on this boundary are easily seen
to be
\beq
\gamma_{ij} = g_{ij} - \ell^2\partial_iF\partial_jF , \qquad
n^a = \left(1-\ell^2g^{ij}\partial_iF\partial_jF\right)^{-1/2}
  \left(\frac{1}{\ell},\ \ell g^{ij}\partial_jF\right) .
\label{a11}
\eeq
Using the expansion (\ref{a1}), it is straightforward to show that
\begin{align}
\sqrt{\gamma} &= e^{2\rho}\sqrt{\overset{(0)}g} 
  + \frac{1}{2}\sqrt{\overset{(0)}g}\left( \overset{(2)}g\,{}^i{}_i
  - \ell^2\,\overset{(0)}g{}^{ij}\partial_iF\partial_jF\right) + \dots ,
\nonumber\\
\sqrt{\gamma}K &= \frac{2}{\ell}e^{2\rho}\sqrt{\overset{(0)}g}
  - \ell\sqrt{\overset{(0)}g}\,
  \overset{(0)}g{}^{ij}\partial_iF\partial_jF 
  + \ell\partial_i\left(\sqrt{\overset{(0)}g}\,\overset{(0)}g{}^{ij}
  \partial_jF\right) + \dots ,\label{a12}\\
\sqrt{{}^{(2)}g} &= e^{2\rho}\sqrt{\overset{(0)}g} 
  + \frac{1}{2}\sqrt{\overset{(0)}g}\,\overset{(2)}g\,{}^i{}_i + \dots, 
\nonumber
\end{align}
all evaluated at $\rho=F$.  Further, since ${}^{(3)}R = -6/\ell^2$, the 
``bulk'' term in (\ref{a9}) contributes\footnote{As in the derivation of 
(\ref{a2}), we need not assume the full ``bulk'' field equations; it is 
sufficient to require that the constraints hold.}
\beq
\int^{\rho=F} d\rho \sqrt{{}^{(3)}g}\left({}^{(3)}R + \frac{2}{\ell^2}\right)
  = -\frac{4}{\ell}\int^{\rho=F} d\rho\sqrt{{}^{(2)}g}
  = -\frac{2}{\ell}e^{2\rho}\sqrt{\overset{(0)}g} 
  - \frac{2}{\ell}\rho\sqrt{\overset{(0)}g}\, \overset{(2)}g\,{}^i{}_i 
  + \dots,
\label{a13}
\eeq
again evaluated at $\rho=F$.  The term proportional to $\rho$ in  (\ref{a13}) 
is the logarithmic divergence described in \cite{HennSken}, and in general 
it must be eliminated by an additional counterterm.  In three dimensions, 
though, the field equations (\ref{a2}) imply that the potentially divergent 
contribution to the action vanishes: the integral of the scalar curvature 
of the two-dimensional boundary is a topological invariant, and is zero 
when the boundary has the topology of a cylinder.

Inserting (\ref{a12}) and (\ref{a13}) into the action (\ref{a9}), we 
obtain
\beq
I_{\mathit grav} 
  = -\frac{\ell}{16\pi G}\int_{\partial M}d^2x\sqrt{\overset{(0)}g}
  \left( \overset{(0)}g{}^{ij}\partial_iF\partial_jF 
  + \frac{2}{\ell^2}F\overset{(2)}g\,{}^i{}_i 
  + \frac{1}{\ell}\overset{(2)}g\,{}^i{}_i \right) .
\label{a14}
\eeq
If we now use the definition (\ref{a10}) of $F$ and the field equations
(\ref{a2}) for $\overset{(2)}g\,{}^i{}_i$, this expression becomes 
precisely the Liouville action (\ref{a8}) with $\lambda=0$.  It is
evident that in this derivation, the Liouville field can be interpreted
as a new dynamical degree of freedom, appearing because the diffeomorphism
(\ref{a3}) is not a true gauge invariance at the boundary \cite{Benguria}.
This result is to some extent implicit in \cite{Navarro}, where the 
authors relate the Liouville field to broken diffeomorphism invariance 
at the conformal boundary, and \cite{BCR}, where the authors note that 
asymptotically nontrivial diffeomorphisms are related to shifts in the 
Liouville field, but neither of these references derives a \emph{dynamical} 
boundary action.  The dynamics \emph{is} derived by Manvelyan et al.\ 
\cite{Manvelyan}, using an approach similar to the present construction, 
and can also be obtained, in its nonlocal Polyakov form, by ``integrating 
the conformal anomaly'' \cite{BCR,Schwimmer}.  In these references, though, 
the emphasis is on the AdS/CFT correspondence; the Liouville field is 
regarded as dual to bulk degrees of freedom, and the aim is to demonstrate 
that the symmetries---including the anomalies and the finite cocycles---match
between the bulk and boundary.

We now have a new interpretation available: the parameter $\varphi$ 
appearing in the asymptotically nontrivial ``bulk'' diffeomorphism (\ref{a3}) 
has become a dynamical variable at the conformal boundary, with dynamics 
given by the Liouville action.  Just as in the Chern-Simons model, 
``would-be gauge transformations'' at the boundary provide new degrees 
of freedom.  Expressed in a slightly different way, the presence of the 
boundary breaks the bulk gauge invariance---in particular, the constraints 
become second class \cite{Benguria,Fjel}---and one can view the Liouville 
field as an analog of a Goldstone boson arising from this symmetry-breaking.

The coefficient in front of the action (\ref{a8}) corresponds to a classical 
central charge \cite{Seiberg} of 
\beq
c = \frac{3\ell}{2G} .
\label{a15}
\eeq
Moreover, the BTZ black hole in Fefferman-Graham-type coordinates has metric
components \cite{Navarro,Ban}
\beq
\overset{(2)}g_{\pm\pm} = \frac{r_+\pm r_-}{4} ,
\label{a16}
\eeq
from which one can obtain the classical value of the Liouville stress-energy 
tensor (\ref{a7}) and thus the classical Virasoro charges.  The resulting 
values of $L_0$, ${\bar L}_0$, and $c$ are precisely those required to obtain 
the correct BTZ black hole entropy from the Cardy formula \cite{Strominger}.   
Recent work by Chen has suggested that the ``nonnormalizable states'' of 
Liouville theory can be counted, and can correctly match the Cardy formula 
\cite{Chen}.  If this is in fact the case, then the results of this paper 
lend support to the conjecture \cite{Carlipc,Carlipd} that the true microscopic 
degrees of freedom of any black hole are ``would-be diffeomorphisms'' made 
dynamical by boundary conditions.

Finally, it is interesting to compare these results to the Chern-Simons
derivation of Liouville theory in \cite{CHvD,Roomana}.  Note first that
the present derivation, like most metric approaches---for example, 
\cite{SkenSolo,Bautier,Schwimmer}---most naturally yields a Liouville 
action with $\lambda=0$.  The Chern-Simons approach, on the other hand, 
naturally leads to a nonzero value of $\lambda$.  The difference can be 
traced to a difference in boundary conditions.  Rather than fixing the
metric at the boundary, the Chern-Simons derivation fixes certain linear 
combinations of the triad $e_a = e_{\mu a}dx^\mu$ and the spin connection 
$\omega_a = \frac{1}{2}\epsilon_{abc}\omega_\mu{}^{bc}dx^\mu$.  This change
has two consequences \cite{BanMend}: the coefficient of the extrinsic 
curvature term in the action (\ref{a9}) is altered, and the extrinsic curvature 
is itself replaced by a first-order expression,
\beq
\int d^2x\sqrt{\gamma}{\tilde K} = \int\omega_a\wedge e^a .
\label{a17}
\eeq
One can always find a local Lorentz frame in which $\tilde K = K$.  But
$\tilde K$ is not invariant under local Lorentz transformations, and new 
degrees of freedom---``would-be local Lorentz transformations''---will now 
appear.  

The $\lambda$ term in \cite{CHvD,Roomana} can ultimately be traced to these 
new degrees of freedom, which are integrated out to give a Liouville potential.  
A direct comparison with this paper is difficult, since the Chern-Simons 
derivation involves field redefinitions that do not easily translate into a 
metric formalism.  But an analysis of the boundary term (\ref{a17}) is actually 
sufficient.  Under a local Lorentz transformation, parametrized as in \cite{CHvD} 
by
\beq
g = \left(\begin{array}{cc}1&x\\0&1\end{array}\right)
    \left(\begin{array}{cc}e^{-\varphi/2}&0\\0&e^{\varphi/2}\end{array}\right)
    \left(\begin{array}{cc}1&0\\y&1\end{array}\right) ,
\label{a18}
\eeq
this boundary term becomes
\beq
\begin{split}
2\int\mathop{Tr}\bigl[(g^{-1}eg)&\wedge(g^{-1}\omega g + g^{-1}dg)\bigr]\\
  &= \int\omega_a\wedge e^a + \frac{\ell}{2}\int dzd{\bar z}\,e^\rho\!\left[
  -e^{\varphi/2}{\bar\partial}y - e^{-\varphi/2}\partial(xe^{\varphi/2})
  +x^2e^{\varphi/2}\partial y\right] .
\end{split}
\eeq
At a surface determined by boundary conditions (\ref{a10}), the equations of 
motion for $x$ imply that $\partial y=0$; a field redefinition $\varphi%
\rightarrow\varphi - \ln(\lambda^{-1}{\bar\partial}y)$, which does not change 
the kinetic or curvature terms in the Liouville action, then introduces exactly
the desired potential.

While this result suggests that the metric and Chern-Simons formalisms are
inequivalent, the difference is not yet fully understood.  It has been
suggested in \cite{BCR} that the Liouville potential term may appear as 
a finite renormalization counterterm.  This is probably not the case in the
holographic renormalization program \cite{Skenderisb},\footnote{Even here, 
there is a delicate issue.  In a conformally flat manifold, only the sign 
of $\lambda$ has physical significance: the value of $\lambda$ may be
changed arbitrarily by a constant shift in $\varphi$.  Thus a potential
term appearing at any order may be physically important.} but the Euclidean 
approach of \cite{Krasnova} suggests that such a term may be unavoidable 
for more complicated configurations in which the boundary is not conformally 
flat.

Regardless of the significance of the potential term, though, the comparison 
with the Chern-Simons approach also demonstrates that in a more general 
setting, the ``would-be diffeomorphism degrees of freedom'' may have 
interesting interactions with other boundary fields.  Results from the 
AdS/CFT correspondence \cite{Emparan} suggest that such couplings are 
necessary for a dynamical description of the BTZ black hole, and may 
ultimately explain such phenomena as Hawking radiation.

\vspace{1.5ex}
\begin{flushleft}
\large\bf Acknowledgments
\end{flushleft}

I would like to thank Nemanja Kaloper and John Terning for suggesting the analogy 
with the Goldstone mechanism, and Kostas Skenderis for valuable comments. 
This work was supported in part by Department of Energy grant DE-FG02-91ER40674.


\begin{thebibliography}{99}
\bibitem{BTZ} M.\ Ba\~nados, C.\ Teitelboim, and J.\ Zanelli, Phys.\ Rev.\ Lett.\ 
 69 (1992) 1849, hep-th/9204099.
\bibitem{Carlipa} S.\ Carlip, Living Rev.\ Relativity 8 (2005) 1, gr-qc/0409039.
\bibitem{Achucarro} A.\ Ach\'ucarro and P.\ K.\ Townsend, Phys.\ Lett.\
 B180 (1986) 89.
\bibitem{Wittena} E.\ Witten, Nucl.\ Phys.\ B311 (1988) 46.
\bibitem{EMSS} S.\ Elitzur, G.\ W.\ Moore, A.\ Schwimmer, and N.\ Seiberg,
 Nucl.\ Phys.\ B326 (1989) 108.
\bibitem{Carlipb} S.\ Carlip, Nucl.\ Phys.\ B362 (1991) 111.
\bibitem{CHvD} O.\ Coussaert, M.\ Henneaux, and P.\ van Driel, Class.\ Quant.\ Grav.\
 12 (1995) 2961, gr-qc/9506019.
\bibitem{Roomana} M.\ Rooman and Ph.\ Spindel, Nucl.\ Phys.\ B594 (2001) 329,
 hep-th/0008147.
\bibitem{Carlipc} S.\ Carlip, in \emph{Constrained Dynamics and
 Quantum Gravity 1996}, edited by V.\ de Alfaro et al., Nucl.\ 
 Phys.\ Proc.\ Suppl.\ 57 (1997) 8, gr-qc/9702017.
\bibitem{BrownHenn} J.\ D.\ Brown and M.\ Henneaux, Commun.\ Math.\ Phys.\
 104 (1986) 207.
\bibitem{Strominger} A.\ Strominger, J.\ High Energy Phys.\ 9802 (1998) 
 009, hep-th/9712251.
\bibitem{Navarro} J.\ Navarro-Salas and P.\ Navarro, Phys.\ Lett.\ B439 
 (1998) 262, hep-th/9807019.
\bibitem{Nakatsu} T.\ Nakatsu, H.\ Umetsu, and N.\ Yokoi, Prog.\ Theor.\
 Phys.\ 102 (1999) 867, hep-th/9903259.
\bibitem{SkenSolo} K.\ Skenderis and S.\ N.\ Solodukhin, Phys.\ Lett.\ B472
 (2000) 316, hep-th/9910023.
\bibitem{Bautier} K.\ Bautier, F.\ Englert, M.\ Rooman, and Ph.\ Spindel,
 Phys.\ Lett.\ B479 (2000) 291, hep-th/0002156.
\bibitem{Roomanb} M.\ Rooman and Ph.\ Spindel, Class.\ Quant.\ Grav.\ 18
 (2001) 2117, gr-qc/0011005.
\bibitem{Krasnova} K.\ Krasnov, Adv.\ Theor.\ Math.\ Phys.\ 4 (2000) 929,
 hep-th/0005106.
\bibitem{Krasnovb} K.\ Krasnov, Class.\ Quant.\ Grav.\ 20 (2003) 4015,
 hep-th/0109198.
\bibitem{Manvelyan} R.\ Manvelyan, R.\ Mkrtchyan, and H.\ J.\ W.\ M{\"u}ller-Kirsten, 
 Phys.\ Lett.\ B509 (2001) 143, hep-th/0103082.
\bibitem{FG} C.\ Fefferman and C.\ R.\ Graham, Ast{\'e}risque, hors s{\'e}rie 
 (1985) 95.
\bibitem{HennSken} M.\ Henningson and K.\ Skenderis, JHEP 9807 (1998) 023,
 hep-th/9806087.
\bibitem{Skenderis} K.\ Skenderis, Int.\ J.\ Mod.\ Phys.\ A16 (2001) 740,
 hep-th/0010138.
\bibitem{BalKraus} V.\ Balasubramanian and P.\ Kraus, Commun.\ Math.\ Phys.\
 208 (1999) 413, hep-th/9902121.
\bibitem{Benguria} R.\ Benguria, P.\ Cordero, and C.\ Teitelboim, Nucl.\ Phys.\
 B122 (1977) 61.
\bibitem{BCR}  M.\ Banados, O.\ Chandia, and A.\ Ritz, Phys.\ Rev.\ D65
 (2002) 126008, hep-th/0203021.
\bibitem{Schwimmer} A.\ Schwimmer and S.\ Theisen, JHEP 0008 (2000) 032,
 hep-th/0008082.
\bibitem{Fjel} J.\ Fjelstad and S.\ Hwang, Phys.\ Lett.\ B466 (1999) 227,
 hep-th/9906123.
\bibitem{Seiberg} N.\ Seiberg, Prog.\ Theor.\ Phys.\ Suppl.\ 102 (1990) 319.
\bibitem{Ban} M.\ Ba{\~n}ados, in \emph{Trends in theoretical physics II},
 edited by H.\ Falomir, R.\ E.\ Gamboa Sarav{\'\i}, and F.\ A.\ Schaposnik
 (American Institute of Physics, 1999), hep-th/9901148.
\bibitem{Chen} Y.-J. Chen, Class.\ Quant.\ Grav.\ 21 (2004) 1153, hep-th/0310234.
\bibitem{Carlipd} S.\ Carlip, eprint hep-th/0408123, to appear in Class.\ Quant.\ 
 Grav.
\bibitem{BanMend} M.\ Ba{\~n}ados and F.\ M{\'e}ndez, Phys.\ Rev.\ D58 (1998) 
 104014, hep-th/9806065.
\bibitem{Skenderisb} K.\ Skenderis, personal communication.
\bibitem{Emparan} R.\ Emparan and I.\ Sachs, Phys.\ Rev.\ Lett. 81 (1998) 2408,
 hep-th/9806122.


\end{thebibliography}
\end{document}